# Computational Courtship: Understanding the Evolution of Online Dating through Large-scale Data Analysis

Rachel Dinh[1], Patrick Gildersleve[1], Chris Blex[1,2], and Taha Yasseri[1,2]*

[1]Oxford Internet Institute, University of Oxford, Oxford, UK
[2]Alan Turing Institute for Data Science and Artificial Intelligence, London, UK

*corresponding author: taha.yasseri@oii.ox.ac.uk

## Abstract

Have we become more tolerant of dating people of different social backgrounds compared to ten years ago? Has the rise of online dating exacerbated or alleviated gender inequalities in modern courtship? Are the most attractive people on these platforms necessarily the most successful? In this work, we examine the mate preferences and communication patterns of male and female users of the online dating site eHarmony over the past decade to identify how attitudes and behaviors have changed over this time period. While other studies have investigated disparities in user behavior between male and female users, this study is unique in its longitudinal approach. Specifically, we analyze how men and women differ in their preferences for certain traits in potential partners and how those preferences have changed over time. The second line of inquiry investigates to what extent physical attractiveness determines the rate of messages a user receives, and how this relationship varies between men and women. Thirdly, we explore whether online dating practices between males and females have become more equal over time or if biases and inequalities have remained constant (or increased). Fourthly, we study the behavioural traits in sending and replying to messages based on one's own experience of receiving messages and being replied to. Finally, we found that similarity between profiles is not a predictor for success except for the number of children and smoking habits. This work could have broader implications for shifting gender norms and social attitudes, reflected in online courtship rituals. Apart from the data-based research, we connect the results to existing theories that concern the role of ICTs in societal change. As searching for love online becomes increasingly common across generations and geographies, these findings may shed light on how people can build relationships through the Internet.

**Keywords:** *Online Dating, Gender, Mating, Communication, Homophily*

## Introduction

As online dating grows in influence as a business and cultural institution, it will become imperative that researchers understand the type of data being collected and the valuable social insights we can glean from the interactions on these platforms. This paper presents a longitudinal study of online dating over a ten-year period, using statistical methods to uncover changes in mate preferences and communication patterns between men and women over time.

A relatively recent phenomenon, online dating is becoming an increasingly relevant site of investigation spanning disciplines as varied as sociology, economics, evolutionary biology, and anthropology (Ellison, Heino,& Gibbs, 2006). Foundational work on mate preferences in online dating, matching markets, and the role of physical attractiveness in online dating has been done by Finkel et al. (2012), Hirstch et al. (2010), and Fiore et al. (2010). Zheng and Yasseri (2016) explored the latent asymmetries in messaging between men and women on these platforms. Though the aforementioned literature is rich and sets a foundation for robust discussion of online dating, no existing study presents a longitudinal approach to online dating. The contribution of this work is the expansive dataset which encompasses over twelve years of user activity, allowing us to better understand not only how these phenomena of interest work in extraordinary detail, but how they have changed over time.

As the internet rose as a social medium used to facilitate communication, it eventually adapted to specialist functions including online dating sites. Online dating is the practice of using dating sites—made specifically for users to meet each other for the end goal of finding a romantic partner (Finkel et.al, 2012). As Michael Norton put it, "Finding a romantic partner is one of the biggest problems that humans face and the invention of online dating is one of the first times in human history we've seen some innovation" (Harford, 2016). In fact, online dating has emerged as one of the most widely used applications on the Internet. Online dating has an annual growth rate of 70% in the United States (Kaufman, 2012). It has also developed into a highly profitable business with growing numbers of people worldwide willing to pay for access to services that will find them a romantic partner. Online dating is now a $2.1 billion business in the US and is expected to continue growing in the foreseeable future (Ortega and Hergovich, 2017). Considering three-quarters of US singles have tried dating sites and up to a third of newly married couples originally met online Ansari & Klinenberg (2015), online dating seems to have shed its old stigma, ostensibly here to stay as the new normal.

When considering online dating, it may be useful to think of these platforms and marriage in general as markets (Roth, 2015). As economist Alvin Roth explains in his book Who Gets What and Why, there can be thick and thin matching markets where thick markets have lots of buyers and sellers (single people in this case) and little differentiation, while thin markets have fewer buyers and sellers and considerable differentiation (Roth, 2015.) For instance, we can imagine that there was a thick market for marrying your high school sweethearts before women started going to college. But as more and more women decided to pursue higher education and enter the workforce, the market shifted to a wider selection of potential spouses for each side of the market.

The increased variety of potential mates gave way to dating phenomena like speed dating, which was a pre-internet predecessor to any modern app with a market design where singles meet many people very quickly, indicate who they're interested in, and only receive each other's contact information if there is mutual interest. But with the rise of the Internet, there is now a thick market for finding love online. More specifically, we can think of these internet-based dating platforms as two-sided matching markets (if we exclude niche platforms for polyamory and non-traditional relationships). This means that there are two sides of the market to be matched, participants on both sides care about to whom they are matched, and money cannot be used to determine the assignment (Azevedo & Leshno, 2016). This model includes high-end management consulting firms competing for college graduates that must attract candidates who also choose them, home buyers and sellers, and many more important markets. Two-sided matching markets have been extensively studied, with the literature splitting them into two categories: the "marriage" model and the "college admissions" model (Azevedo & Leshno, 2016).

Becker's (1973) marriage model assumes simple preferences, with men and women ranked vertically from best to worst. This model and its assumptions have been applied to diverse problems such as explaining gender differences in educational attainment, changes in chief executive officer wages, and the relationship between the distribution of talent and international trade (Grossman 2004; Gabaix and Landier 2008; Terviö 2008; Chiappori, Iyigun, and Weiss 2009; Galichon, Kominers, and Weber 2016; Bojilov and Galichon, 2016).

Another line of research follows Gale and Shapley's (1962) college admissions model which allows for complex heterogeneous preferences. This model is a cornerstone of market design and has been applied to the study and design of market clearing houses such as matching residents to hospitals and students to charter schools. This begs the question: who gets matched with whom in the online dating matching market? Are differences in dimensions of type mostly horizontal (e.g., some pairs make better matches than others, following the college admissions model), or vertical (e.g., there are some people that we can universally agree are more desirable mates than others, following the marriage model)?

There are "superstar" users who attract lots of attention and matches on any given platform. In some cases, the top 5 percent of all men on a platform receives twice as many messages as the next 5 percent and several times as many messages as all the other men (Oyer, 2014). But it would be incorrect to assume these superstars would be universally appealing to all users and that popularity alone determines matches. Instead, it could be useful to consider the economic concept of assortative mating observed in offline marriage markets, and how online matching reflects or deviates from this behavior.

Positive assortative mating or matching occurs when people choose mates with similar characteristics. Empirical evidence strongly suggests that spouses tend to be similar in a variety of characteristics, including age, education, race, religion, physical characteristics and personality traits (Becker, 1991; Dupuy and Galichon, 2012; Orece and Quintana-Domeque, 2010; Qian, 1998; Silventoinen et al., 2003; Weiss and Willis, 1997). This phenomenon can be measured and observed in online dating markets when we inspect the pairs. Using data from an online dating site, Hirsch et al. found that although physical attractiveness and income are largely vertical attributes, preferences concerning a partner's age, education, race, and height tend to sort assortatively. Likewise, the examination of "bounding" characteristics shows that life course attributes, including marital status, whether one wants children, and how many children one has already, are much more likely than chance to be the same across the two users in a dyadic interaction (Fiore, 2002).

In other words, mate preferences are not simply vertical, meaning we always want mates with the highest level of education, income, etc. Rather, horizontal preferences and preferences for similarity in particular, play an important role (Hitsch, Hortaçsu, & Ariely, 2010). Overall, users with similar education levels are three times as likely to match. As we can observe, assortative mating occurs in both online and offline contexts and can partially help explain why these markets still tend to be efficient.

Newer niche dating apps that only admit users from certain echelons of society may be changing the way we sort and actually exacerbate existing assortative tendencies. A recent *Bloomberg* report argues that dating apps, particularly elite ones like the *League* and *Luxy*, may be worsening economic inequality by making it easier for couples to pair by socioeconomic status. The League famously only admits graduates from top universities, while Luxy purports that the median income of users on its platform is $500,000. Instead of meeting someone at a bar or other social setting, singles can now use apps to find their economic and educational equivalent. While one might argue that this phenomenon already occurs offline, according to Bloomberg, "these services help facilitate unions between educated, affluent Millennials who are clustering in such cities as San Francisco and New York" — indirectly intensifying economic inequality.

While those may be exceptional cases, some combination of an individual's attributes and potential partners' preferences dictate market dynamics both in online and offline contexts. This means that an individual may have high desirability for one person and low desirability for another, and the preferences may not necessarily be monotonically related to their attributes. Efficient matching in this market thus relies on the existence of pairs of mutually desirable agents in a setting where preferences are heterogeneously distributed. As Hitsch, et al. note, these markets tend to naturally resolve into pairs of mutual desirability (Hirsch et. al, 2010). This might seem somewhat obvious, but is remarkably observed, measured, and explained in the online dating environment.

Online platforms provide us with a unique opportunity to study the economic and evolutionary concepts of sorting and matching. While part of this is due to the ability to observe and classify user attributes, preferences, and behavior in great detail, it is also due to the unique lack of search frictions in online dating markets. Certainly, a main reason for the existence of online dating sites is to make the search for a partner as easy as possible (Hirsch et. al, 2010). Yet despite the wealth of insight user-generated data online dating has revealed about latent and stated mate preferences, there remains significant uncertainty regarding the way these preferences have evolved over time.

Sociologists often assume that society has become more egalitarian, and that these pluralist ideals have translated into a more equal quest for love (Ferrante, 2007). It would then follow that people's mate preferences have become more pluralist, switching from sorting based on ascribed traits to sorting based on acquired traits. Ascribed characteristics, as used in the social sciences, refers to properties of an individual attained at birth. The individual has very little, if any, control over these characteristics. In other words, based on the progress we have reportedly seen over the past decade in social integration, we would expect to observe users placing less importance on inherited traits like ethnicity and height, and more importance placed on characteristics achieved through merit such as education.

**RQ1:** This research explores how stated and revealed mate preferences have evolved over the last decade and whether the claims of a more egalitarian society are in fact reflected in online dating and mate selection.

In mate selection and especially in online dating, there seems to be a preoccupation with physical beauty (Rhodes, Simmons, & Peters, 2005)). Historically, theories of interpersonal attraction and interpersonal judgments have emphasized the importance of physical attributes over other factors such as personality and intelligence (e.g., Dion et al., 1972; Walster et al., 1966). Accordingly, online dating sites often urge their users to post photos of themselves to increase the chances that potential dates will contact them. Dating services like Grindr and Tinder have gone even further by doing away with detailed profile descriptions altogether, allowing users to base their dating decisions on physical appearance alone (Hannah Fry, 2015). Indeed, 85% of interviewees in a study of Australian online dating users said they would not contact someone without a photo on his or her profile (Whitty and Carr, 2006).

Only a few studies so far have considered how users judge attractiveness online generally or in online dating in particular and how this translates into messaging strategy. Ellison describes the strategies employed by online dating users to interpret the self-presentations of others. Primarily, the participants they interviewed made substantial inferences from small cues, lending support to Walther's theory of Social Information Processing (Walther, 1992). For example, one woman felt that people who were sitting down in their online dating profile photos were trying to disguise that they were overweight (Ellison et al. 2006). Fiore found that in line with past research on the psychology of attraction, the attractiveness of the photograph were the strongest predictors of whole profile attractiveness in online dating (Fiore, 2008).

But while it is evident that the attractiveness of one's photo is important in determining overall perceived attractiveness of an online dating profile as a whole, predicting popularity based on looks alone is much more ambiguous. Christian Rudder explored the importance of attractiveness in online dating and found that how good-looking you are does not dictate how popular you are on an online dating website. In fact, having some people think that you are ugly can work in your favor (Fry, 2015).

To try and test how attractiveness might predict popularity, the OkCupid team took a random sample of 5,000 female users and compared the average attractiveness scores they each received from other users with the number of messages they were sent in a month. They found it is not just the better-looking people who receive lots of messages. Using the spread of attractiveness ratings, they identified people who divide opinion on their attractiveness. These polarizing users ended up being far more popular on internet dating sites than universally attractive people. Users rated 4 out of 5 were penalized, while people at the extremes of the spectrum at 1 and 5 received were much more likely to receive messages. In essence, the most beautiful users will always do well but users whose attractiveness divides opinion are better off than those who everyone agrees is just quite cute.

Fiore and Donath (2005) also explored this question of predicting popularity, but used self-reported attractiveness instead of attractiveness scores given by other users. They found that men received more messages when they were older, more educated, and had higher levels of self-reported attractiveness. Women received more messages when they did not describe themselves as "heavy," had higher levels of self-reported attractiveness, and posted a photo on their profiles.

Among online daters, sending signals such as a "Superlike" or "Smile," or "favoriting" a user can be a way to let them know a user is interested. In a notable study using a Korean dating/marriage site, researchers found evidence that sending a signal increased the total number of dates. But the study was also able to use various measures to determine who were the most sought after people on the website and who were not as sought after, by ranking participants as high, medium, and low in the distribution of rated attractiveness. And it turns out the most popular people on the website were not very responsive to virtual roses (Lee, et al., 2011). Because their attitude was "well, of course, that person's interested in me." Instead, the virtual rose was most effective on the middle desirability group which did not have as many great dating options and was almost twice as likely to accept a proposal sent with the costly signal of a rose.

This brings to light issues with signaling optimization: Despite the positive effect of sending roses, a considerable portion of participants did not use their roses and even those who exhausted their supply did not properly use them to maximize their dating success. It seems there are substantial tradeoffs in preference signaling. Reminiscent of the bar scene with John Nash in A Beautiful Mind, a user could send their signal to the 'blonde' or the most attractive female on the platform, who would be their number one pick. But if everyone uses this strategy, chances of success are low. Instead, users would be better off using their costly signal on a medium quality mate where chances of reciprocity are higher. By the same token, it seems like success could be almost guaranteed by seeking out the least desirable mate and sending a signal, but this is obviously not optimal. So there's a trade-off in choosing who to send a costly signal such as a favorite or message to that goes back to the aforementioned difference in user "quality' or desirability.

**RQ2:** <u>This research will explore the impact of user attractiveness on messaging patterns and whether it is a powerful predictor of "success" in online dating.</u>

In the social sciences, gender is a built-in variable that can account for measurable differences in behavior. (Rakow, 1986). While non-binary users and same-sex dyads are a growing segment of online dating users, the dataset examined in this work consists exclusively of heterosexual dyads. One of the main research areas related to online dating systems is the difference in messaging behavior between men and women on these platforms. But in order to meaningfully investigate computer-mediated communication between genders, it is important to first understand underlying patterns of offline communication between heterosexual dyads that may be reflected, moderated, or exacerbated online.

Examining single women's use of the telephone in heterosexual dating relationships, Sarch found that in line with gender norms at the time of the study, subjects expected men to pursue women (Sarch, 1993). Additionally, on occasions when a woman ever took initiative and started a conversation, she expected her partner to "overcompensate" by reaching out with more frequency. Subjects also reportedly saw the frequency of how often their dates called as an indicator of how well the relationship was going or how often their date was thinking about them.

In keeping with these two indicators, subjects did not want to be perceived as the pursuer so they limited the frequency of their own calls by ensuring that each one was "carefully executed so that sufficient time elapsed between multiple phone calls" (Sarch, 1993, p. 141). This phenomenon has not entirely disappeared—Ansari and Klinenberg observe, "the fear of coming off as desperate or overeager through texting" as a common concern in recent focus groups (2015). Despite coming 22 years after Sarch's study, Ansari and Klinenberg's research (2015) shows that initiator status and contact frequency equating to interest has translated from telephone calls to modern online messaging culture.

Besides the stigma against female initiators, another reason initiators tend to be male has to do with the way incentives are structured in online dating. About 60% of the men in Whitty and Carr's study saw online dating as a "numbers game" (2006). Given the seemingly endless number of profiles available, individuals could keep trying until they get a response, meaning they are not fully interested in some of the profiles they send messages to. Instead, they would send a large number of initiations regardless of actual interest and see which women reciprocate, filtering at the response level.

The result is staggeringly lop-sided activity levels for men and women. Men are on average twice as active as women in online dating apps– skewing an already imbalanced gender ratio; taking into consideration activity level, the gender ratio of the active user base is more like 80:20 (Harford, 2018). Rudder (2014) confirms this, showing that even the most attractive men receive fewer messages than women on average. In turn, since women are often inundated with date requests, they are less compelled to respond to each request (Tong & Walther, 2011). Fiore confirms this, finding that women responded more selectively than men, answering 16% of the time compared to men's 26% reciprocation rate (2010).

Zhang and Yasseri found that messages were five times more likely to have been initiated by a man than by a woman even in dating applications that allow users to communicate only after they have mutually signaled their interest (2015), in line with previous work that found men to be the main initiators in heterosexual conversations (Finkel et al., 2012; Whitty, 2012; Tong & Walther, 2011; Whitty, Baker, & Inman, 2007). Fiore also confirms this, finding that rates of initial contact differed sharply by gender. Men initiated a median 1 contact per day compared with 0.875 for women (Fiore, 2010). Given this difference combined with the greater number of men on the site, women tended to be contacted much more often than men, a median 2 times per day, compared to 0.5 for men. Finally, more popular men and women — those who were contacted more often per day — initiated contact with others slightly less often, confirming economic theory that "high quality" users need not pursue others as actively.

**RQ3:** This work will explore whether the previously established phenomenon of gender asymmetry in online dating messaging behavior has remained stable, lessened, or grown over time.

We integrate the previously mentioned literature on attractiveness and selectivity to investigate how user behaviour and strategy varies across different facets of communication; searching for partners to initiate contact with, and selecting which users to reply to when they have some awareness of their attractiveness or signals of their success. As well as studying variations of behaviour in the population, we are also motivated by research around Dunbar's number (Dunbar, 1992) to study what limits and commonalities might be present in the data around users' communication.

**RQ4:** This research will analyse what general communication patterns exist in eharmony users, as well as how different facets of online daters' success relate to their selectivity.

Referring back to the "college admission" model that suggests strong homophily in seeking partners, most studies have overlooked whether a match based on homophily actually translates into initiation of contact and communication between users in a liquid market and in the absence of search friction. Given the abundance of inactive users and the asymmetry in the activity between male and female users, matching alone is insufficient to determine whether online dating is driven by homophilic tendencies. hence we form our last research question as the following.

**RQ5:** Does similarity between the parties involved in a computationally made match map into initiation of contact and successful communication?

Moreover, homophily is unlikely to be uniformly distributed across all characteristics for all users. For instance, some users will weigh age differences stronger than others. Whilst there seem to be some hints that especially demographic or socioeconomic features play an important role, the exact relationships and relevant variables are still ambiguous.

**RQ6:** Given presence of homophily, which are the decisive dimensions and variables predicting successful communication?

## Data and Methods

To address the aforementioned research questions, this work analyzes a data set obtained through a collaboration with eHarmony UK, a major web-based online dating system. Broadly speaking, web-based online dating systems include the following (Fiore and Donath, 2004):

- Personal profiles for each user, which include demographic and other fixed-choice responses, free-text responses to prompts, and, optionally, one or more photographs.
- Searching and/or matching mechanisms so that users can find potential dates from among the thousands of profiles on a typical system.
- Some means of private communication that permits users to contact potential dates within the closed online dating system without disclosing an email address, phone number, or identifying information. This usually means a private mail system, but it sometimes also includes the ability to send "smiles" or some other token of interest.
- Optionally, other forms of self-description: for example, the results of a personality test, or multimedia uploaded by the user.

eHarmony's platform follows the typical format of other online dating systems, including personal profiles and messaging channels, but is distinctive in that users can only communicate with matches selected through an algorithm. This matching algorithm is based on responses from a questionnaire each user completes upon registration. This work will utilize stated mate preference and demographic data collected through this questionnaire, as well as the user interactions that occur after the match – namely, messaging communications. Since the aim of online dating systems is to facilitate face-to-face contact (Whitty et al., 2007; Finkel et al., 2012), with communications being a prerequisite to any offline encounters, this research will operationalize communications received, sent, and reciprocated as meaningful measures of interest and popularity.

The sample dataset used in this work was generated from users who registered during a randomly selected month (March) for each year between 2007 to 2018. Data was not sampled from January or February since they are probably not the most "typical" months, due to holidays including New Year's Day and Valentine's Day. The data was generated from user profiles and private messaging activity on the dating site over the twelve-year period and consists of 149,440 unique heterosexual users from across the United Kingdom. All demographic information and gender were self-reported upon registration. The dataset did not contain any users identifying as non-binary so the term "gender" in this work will refer to male or female self-identification. Since we sampled for all users registered in the month of March for each year, the registration month for all cases are the same, but the total cases of each year varies as reported in Table 1.

Table 1: Number of users in the dataset for each year

| YEAR | TOTAL CASES | MALES | FEMALES |
|---|---|---|---|
| 2007 | 50 | 40% | 60% |
| 2008 | 133 | 45.11% | 54.89% |
| 2009 | 5,453 | 37.8% | 62.2% |
| 2010 | 30,942 | 47.64% | 52.36% |
| 2011 | 16,439 | 46.23% | 53.77% |
| 2012 | 15,329 | 51.15% | 48.85% |
| 2013 | 14,344 | 47.01% | 52.99% |
| 2014 | 13,967 | 48.16% | 51.84% |
| 2015 | 13,731 | 46.97% | 53.03% |
| 2016 | 14,059 | 49.66% | 50.34% |
| 2017 | 14,863 | 49.03% | 50.97% |
| 2018 | 10,129 | 50.97% | 49.03% |

Table 2 summarizes the variables describing users and their behavior. In the following sections, the variables are grouped by type and defined in further detail.

Table 2: Description of variables in the dataset for each user

| USER-LEVEL DATA |
| --- |
| **USER CHARACTERISTICS** (demographic and profile information) |
| Age |
| Gender |
| Registration date |
| Smoking level (1-7) |
| Drinking level (1-7) |
| Number of photos |
| **MATE PREFERENCES** (Likert scale) |
| Ethnicity |
| Education |
| Income |
| Distance |
| Age |
| Height |
| Smoking level |
| Drinking level |
| **PSYCHOMETRIC QUESTIONNAIRE** (Likert scale) |
| Attractiveness |
| Neuroticism |
| Athleticism |
| Agreeableness |
| Religiosity |
| Sexual |
| Romantic |
| Cleverness |
| Conflict Resolution |
| Altruism |
| Conscientiousness |
| **COMMUNICATIONS DATA** |
| Communication initiations sent |
| Communication initiations received |
| Communication rate (total communication initiations received/total profile views by other matches) |

Mate preferences were collected upon registration through a questionnaire asking about the importance of different match criteria based on a Likert-type scale, ranging from Not Important, Somewhat Important, and Very Important. The variable for user attractiveness used in RQ2 was created using the average score of self-reported responses to the following questions: "How stylish do you consider yourself?" "How attractive do you consider yourself?" and "How sexy do you consider yourself?" on a scale from 1-7. The remaining psychometric variables were also created using a similar formula of questionnaire responses.

Communication-level data was inspected by gender and initiation. Initiation refers to whether the sender of the message is the user who had sent the first message in the conversation. The gender of a given message sender is tied to the initiator of a message, as all messages in the dataset are between heterosexual matches—for example, if the conversation initiator is male, the responder would be female.

When computing messaging statistics, the primary measure was not the sheer number of messages sent or received but the number of distinct people whom a user contacts or is contacted by. This places the focus not on how many messages a pair exchanges, but rather on distinct cases of initiated contact. In particular, one key focus of this study is predicting "popularity" in an online dating system. This study falls in line with Fiore and Donath's (2010) theory that a person's popularity on an online dating site is best indexed by the average number of people who initiate contact with him or her. However, this work deviates from their belief that this measure doubly serves as a reasonable proxy for overall attractiveness as well. While Fiore and Donath assumed that more attractive people on average receive more unsolicited attention than less attractive people, this work seeks to tie in later findings from Rhodes et al. (2005), Rudder (2014) and Fry (2015), and understand user attractiveness and popularity as two distinct variables. Finally, to control for the fact that the number of matches each user has is an artifact of the algorithm which has slightly changed over the years, we sought to normalize the number of contacts received by the number of profile views each user received, creating a new "communication rate" variable for each user. This metric is an approximation which accounts for users who might be much more active and get more site exposure than others.

Further to the user profile data, we also possess data for messages for a subset of supplied users (70,508 users, 1,048,575 matches – which can be acted upon by both people, one person, or neither). This is presented in Table 3. There are 5 distinct ways of establishing contact with another user on eharmony: Closed Ended Question, Fast-Track, Must-Haves / Can't Stand, Open Ended Question, Icebreaker, and Open Communication.

Table 3: Data Parameters

| MATCH-LEVEL DATA |
| --- |
| **USER CHARACTERISTICS** |
| Match ID |
| User 1 ID |
| User 2 ID |
| **FOR EACH FORM OF COMMUNICATION** |
| Initiation Date |
| Initiator Gender |
| Response Date |
| Responder Gender |

*Ethics*

Given that this research used data generated from real users of an online dating platform, privacy was a top ethical concern throughout data collection and analysis. As such, proper precautions were taken to preserve privacy and ensure the anonymity of users. During data collection, the eHarmony team excluded any personally identifiable information such as names, payment information, and address to prevent triangulation. Data was captured, transferred, and stored on a password-protected computer with an encrypted hard-drive. Users are only identified by an anonymous user ID number.

Confidentiality and data transfer agreements were signed both by the company and the University of Oxford to preserve privacy rights for eHarmony users. Users are informed of data collection and analysis efforts at the time of sign up, when presented with Terms & Conditions and Privacy Policy agreements. eHarmony UK operates fully certified under the EU-US and Swiss-US Privacy Shield frameworks. The University of Oxford CUREC (Central University Research Ethics Committee) approved all handling of data and research methods. The CUREC number for this project is SSH OII C1A 18 032.

# Results
*Partner Preferences*

The majority of users in our dataset are from London, followed by Manchester, Birmingham, Glasgow, and Bristol. The minimum age is 18 and the maximum age is 98. The mean age of the users in the sample is 38 while the median age is 37. The gender makeup of the dataset is 52% female and 48% male. Most users are non-religious (53%), followed by Christian (34%), then Other and Muslims. Most users have never been married (67%), 24% are divorcees, and 3% widowed. All users are engaging in heterosexual interactions on the platform.

We inspected the stated level of importance for both men and women in regards to six different mate preference criteria: income, education, age, religion, smoking level, and drinking level.

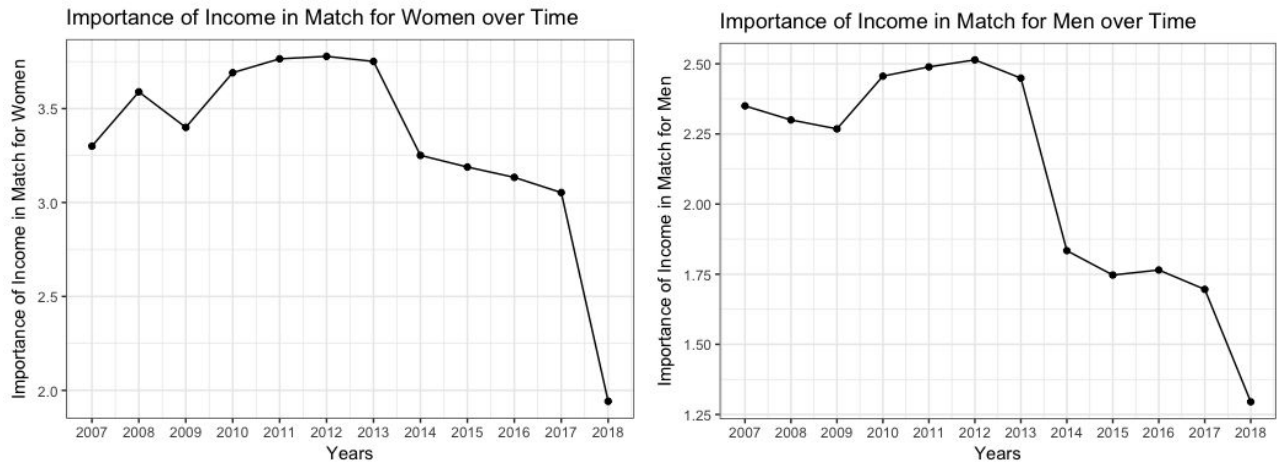

**Figure 1:** Importance of income of the partner (2007-2018).

Regarding the average importance of income (Figure 1), women have a consistently higher mean than men, meaning that they consider income of a potential match more important than men do for all years 2007-2018. This difference between female and male preference for income is statistically significant (at $p<.05$). Nevertheless, for both genders, after a Post-Financial Crisis increase, we see that the importance of the income of the partner has been decreasing over more recent years.

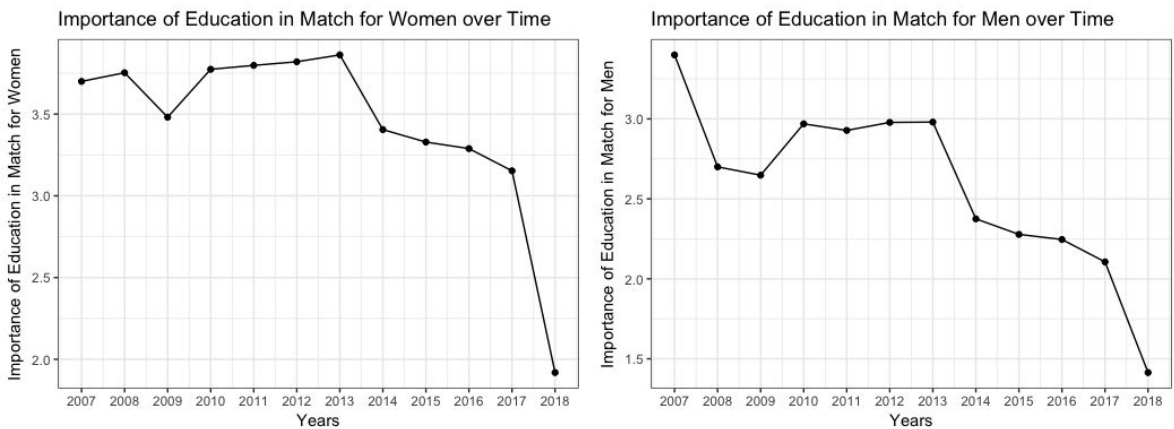

**Figure 2:** Importance of education level of the partner (2007-2018).

As for education (Figure 2), women have a consistently higher mean than men, meaning they consider education of a potential match more important than men do for all years 2008-2018. The overall trend is very similar to the one for income: an increase around 2010-2013 and then a steady decrease.

When it comes to age (Figure 3) too, women have a higher mean than men for all years 2008-2018, meaning they consider age more important than men do. Change in average score over time is not monotonic for men or women.

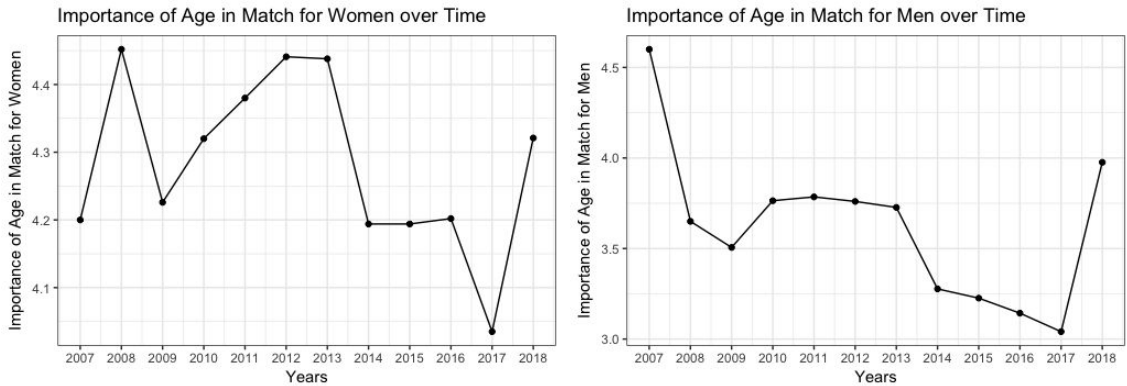

**Figure 3:** Importance of age of the partner (2007-2018).

Regarding the importance of smoking and drinking levels, there is no clear pattern in the changes of "average" over time, mostly because users are polarized by prospective partners smoking (58% Not Important, 40% Very Important). Users are less concerned with prospective partners drinking alcohol (77% Not important to Somewhat Not Important) in 2018.

Finally, comparing all online dating mate preferences with each other, women consider these traits most to least important: smoking level, ethnicity, drinking level, education, income, and then religion. For men, the order is: smoking level, ethnicity, drinking level, religion, education, and finally income.

*Physical Attractiveness*

The second research question we sought to answer was to what extent self-perceived physical attractiveness determines popularity in online dating. When investigating the relationship between self-perceived physical attractiveness and *communication rate* (communication initiations received over profile views), we found no significant change from year to year. Thus, here the results will be discussed in context of the overall dataset and not in respect to change over time. Looking at the aggregate dataset, users do tend to have a higher communication rate as self-perceived attractiveness increases, but the rate of increase appears to first plateau and then decreases.

When the data is separated by gender (Figure 4), the pattern holds for both men and women, although the slope for women between 2 and 6 attractiveness (0.023) is significantly larger than for men (0.008).

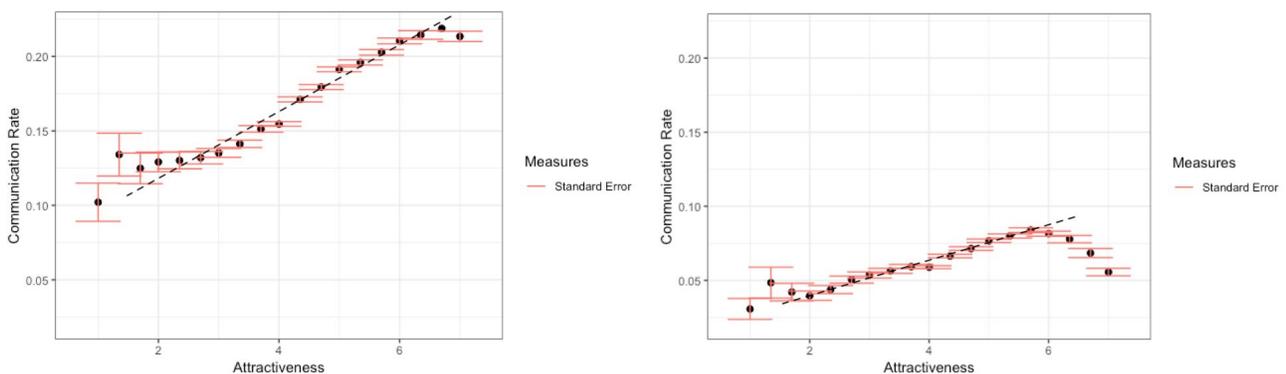

**Figure 4:** Communication rate vs attractiveness. The slope of the linear fit is 0.023 for women (left) and 0.008 for male (right) profiles.

An intriguing observation here is that for both genders, the most "successful" profiles based on the chance of receiving a message from a visitor are not the ones whose owners have ranked themselves the most attractive. For male profiles this is even more evident.

*Communication patterns*

Next, we addressed the third research question regarding the asymmetries in communication initiation between men and women in online dating (Figure 5). The *initiator ratio*, or percentage of sent communication initiations over total communications for the average female user trends downward over time. For men, the initiator ratio shows the opposite trend, increasing from 2008 to 2013, with a small dip in 2014 before climbing again. Initiator ratio drops for both men and women in 2018. It should be noted that percentages for men and women do not add up to 100% for each year because the calculation is not for total communications sent and received between men and women within a single year, but for the lifetime of each user profile sampled from March of each given year. As evidenced, men on average consistently initiate more communications than women. The difference between men and women's average initiator ratios is persistent from 2009-2018.

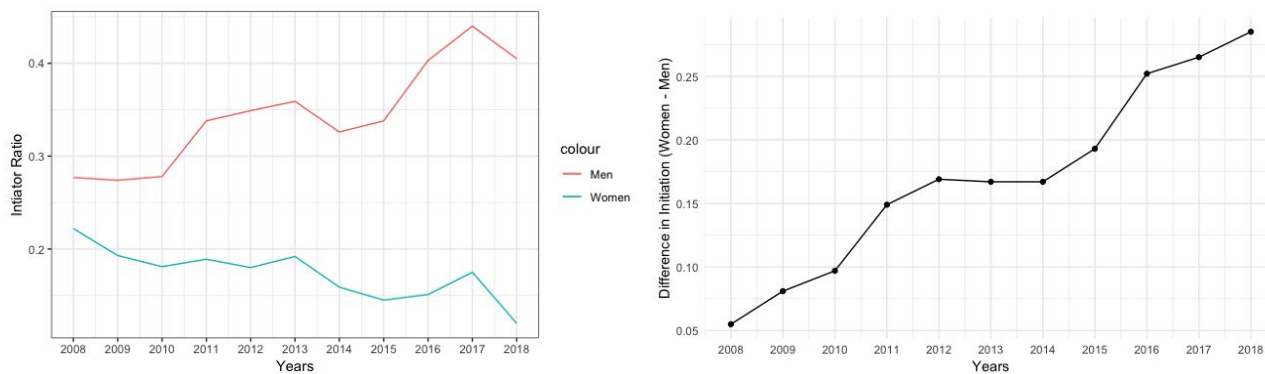

**Figure 5:** Initiation ratios for men and women over time (left) and difference in initiation ratios between men and women over time (right).

We look further to the individuals in the matching data to investigate initiation ($I$) and reply rate ($R$) in time, defined as the average number of requests initiated or replied to by a user per day. Messages are binned into individual days, and users tend to go through phases of high and low activity, presumed to be when they might be 'exclusive' with a partner, temporarily unsubscribed from the platform, or simply too busy or uninterested in engaging in online dating. As such, we take the average of the inverse time between days of activity, weighted by the number of interactions on the days in question.

There is a sizable number of users who seemingly are active at the very start of their subscription, then not at all until the end of their subscription, we speculate this is due to a reminder email encouraging one last stab at finding romance. This results in an artificial peak of users whose activity is around 1 request per 365 days. These users are removed from the following analysis, so as to focus on those who regularly use the site. We also remove users who are only active on one day. The cumulative distributions are shown in Figure 6. along with the 50%, 90%, 95%, and 99% quantiles. The quantiles are given as number per day, so initiation rate at 95% at 1.735 per day translates to 12.15 new initiations per week. Similarly reply rate at 95% at 0.943 per day translates to 6.60 new replies per week.

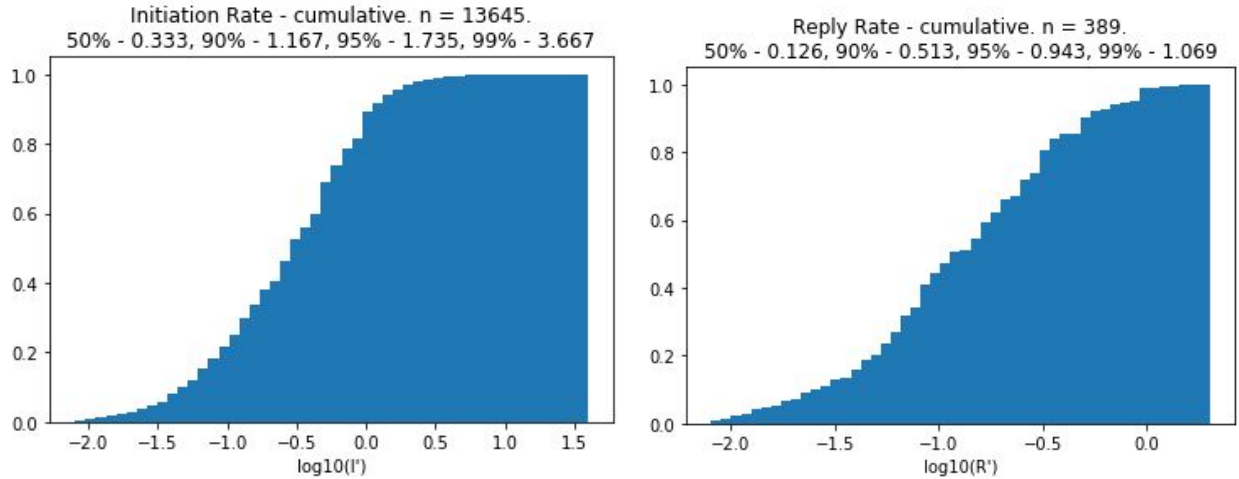

**Figure 6:** Cumulative histograms for Initiation Rate and Reply Rate

The match level data offers the opportunity to investigate the properties of attractiveness and selectivity of users when they send, but also receive requests – not necessarily symmetrical. The match-level data is aggregated by user, allowing us to study the range of behaviours of users when sending or receiving requests, as displayed in Figure 7.

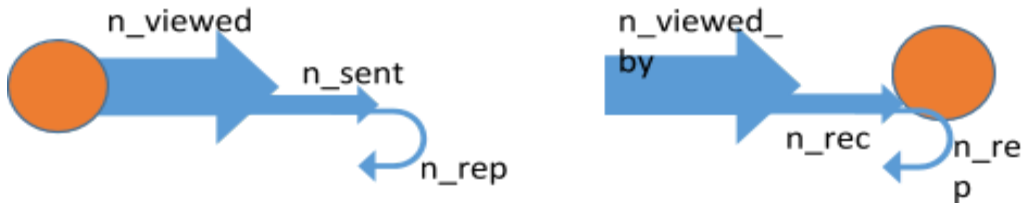

**Figure 7:** Schematic for the fraction of possible partners that a user views, sends requests to and gets replies from, as well as for the fraction of possible partners a user is viewed by, receives requests from, and whose requests are replied to.

We define the following measures for *communication attractiveness when sending* and *receiving* requests;

$$A_s = \frac{n_{rep}}{n_{sent}} \qquad A_r = \frac{n_{rec}}{n_{vb}}$$

Note that this defines attractiveness according to the success of a user, rather than self-perceived physical attractiveness, as used previously. We also calculate the *initiation ratio* and *reply ratio*, as used previously, how likely a user is to engage with another user when presented with a possible match

$$I = \frac{n_{sent}}{n_v} \qquad R = \frac{n_{rep}}{n_{rec}}.$$

Selectivity when sending and receiving requests, or how picky a user is when selecting partners, is consequently defined as;

$$S_s = 1 - I \qquad S_r = 1 - R,$$

where S~0 indicates users send/reply to requests indiscriminately, and S~1 indicates users send/reply to a very small fraction of possible requests. Note that it is also possible to calculate $A_r$ and $I$ from the user data.

$A_r$ = 'COMM_RATE' and $I$='COMM_SENT'/'MATCHES_VIEWED'. We also refer to $I$ as 'SEL_RATE' when it is calculated from the user profile data. Thresholds for the studied users are applied such that;

- Each user has more than 5 potential interactions for the attractiveness / selectivity measures of study e.g. n_viewed or n_viewed_by – to remove effects of observations with small sample size.
- At least 1 interaction is undertaken by the user e.g. n_sent, n_rec, n_rep – to remove inactive users.

We study the correlation of the attractiveness and selectivity features with each other, plotted in Figure 8.

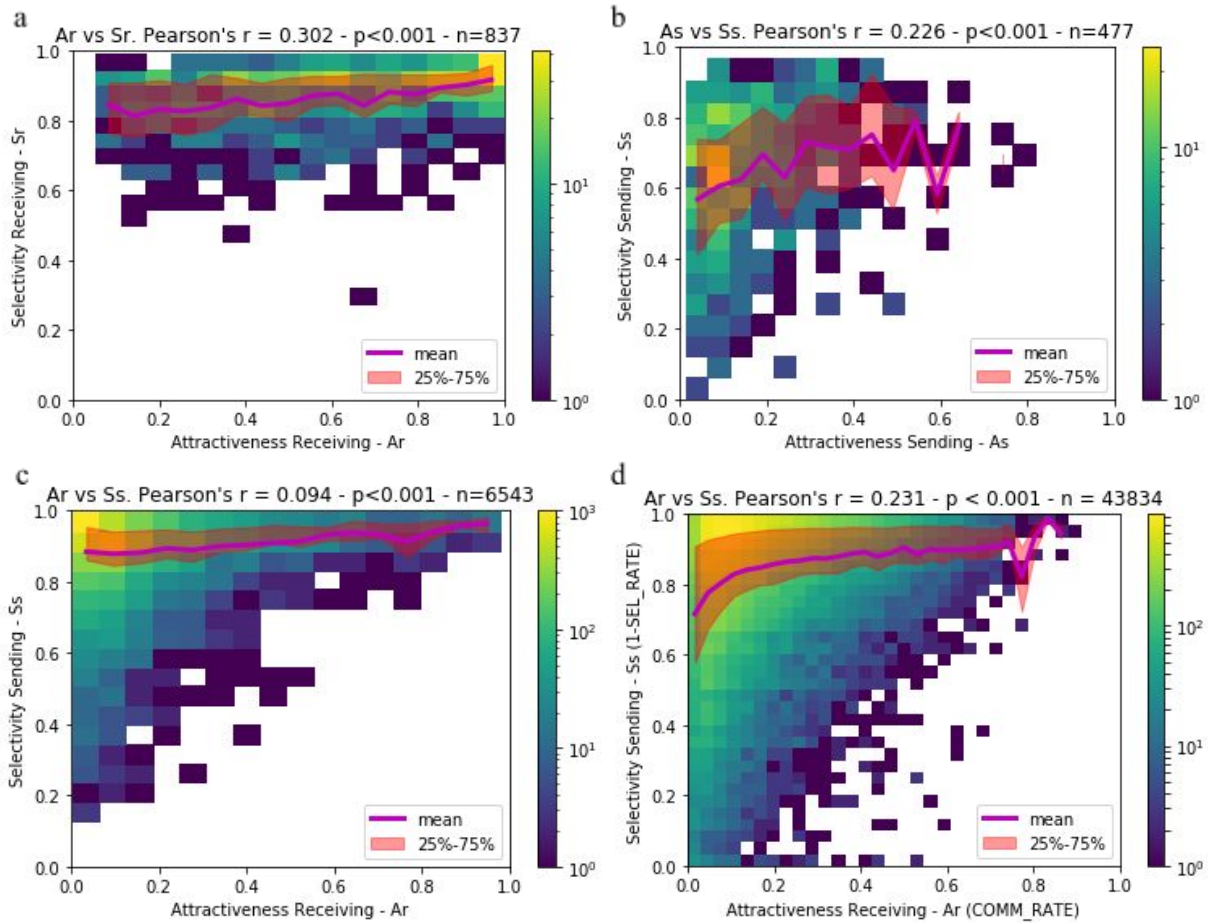

**Figure 8:** Relationships between attractiveness and selectivity for both receiving and sending

We observe a weak, positive, significant correlation between attractiveness and selectivity for all 4 of these figures. This is as expected for $A_r$ vs $S_r$ in Figure 8(a), where those who receive lots of requests are able to select their preferred partners, whereas those who receive fewer matches are forced to be less choosy. This is less trivial for Figure 8(b), $A_s$ vs $S_s$, though users act on feedback based on how many replies they receive to adjust their expectations, as well as self-perception of attractiveness. Finally, Figures 8(c) and 8(d), $A_r$ vs $S_r$ and 'COMM_RATE' vs $1 -$ 'SEL_RATE' (when calculated from the user profile data), shows the correlation between attractiveness when receiving requests, and selectivity when sending requests, one step removed from each other, and reliant on self perception of attractiveness compared to potential matches.
More attractive people, whether initiating or receiving contact, are more selective with the fraction of people that they message.

## *Personality features*

Finally, we built a multiivariate regression model to determine which variables could predict "success" in online, as measured by communication rate (communication initiations received over profile views by matches). After transforming skewed variables to normalize distribution and standardizing all coefficients, the results of the multivariate regression against communication rate are reported in Table 4.

**Table 4.** Multvariate linear regression models to determine which variables predict receiving communication initiations. The first model is for men receiving communication initiations from women, and the second is for women receiving communication initiations from men.

| Variable | Men | | Women | |
|---|---|---|---|---|
|  | Estimate | S.E | Estimate | S.E |
| *(Intercept)* | *7.917e-04* | *7.710e-03* | *0.0007* | *0.005* |
| User Drinking Level | 6.070e-02*** | 8.003e-03 | -0.010 | 0.006 |
| Age | -5.067e-02*** | 8.112e-03 | -0.329*** | 0.006 |
| Communications Sent | -2.564e-01*** | 8.564e-03 | -0.174*** | 0.006 |
| Total Photos | 1.626e+02*** | 8.658e+00 | 0.151*** | 0.006 |
| Neuroticism | -2.433e-02** | 8.038e-03 | 0.008 | 0.006 |
| Athleticism | 1.760e-01*** | 8.604e-03 | 0.256*** | 0.006 |
| Agreeableness | 2.999e-02** | 1.157e-02 | -0.018* | 0.008 |
| Sexual | -1.165e-01*** | 8.857e-03 | -0.106*** | 0.006 |
| Romantic | 3.700e-02*** | 1.060e-02 | 0.093*** | 0.007 |
| Cleverness | -3.300e-02*** | 8.465e-03 | 0.0002 | 0.006 |
| Conflict Resolution | -6.110e-02*** | 9.201e-03 | -0.038*** | 0.006 |
| Altruism | 6.023e-02*** | 1.044e-02 | 0.045*** | 0.007 |
| Conscientiousness | -1.215e-02 | 7.998e-03 | 0.006 | 0.006 |
| *Number of observations* | 71674 | | 77765 | |
| *Multiple R-squared* | 0.134 | | 0.256 | |
| *Adjusted R-squared* | 0.133 | | 0.256 | |
| *Degrees of Freedom* | 14570 | | 21649 | |

*** $p < 0.001$    ** $p < 0.01$  * $p < 0.05$

The results of the individual models for each gender reveal that there are different variables that predict success for men and women. Since the coefficients are standardized, we can compare between variables within each gender. For men, being altruistic and having a higher drinking level were the strongest predictors of receiving messages, while being older and more oriented toward conflict resolution were the most negative predictors of receiving messages. For women, cleverness, neuroticism, and drinking level had no impact on predicting likelihood of receiving messages. Being older was the strongest negative predictor of receiving messages, while being athletic was the strongest positive predictor. Similar to the results for men, sending communications and being sexual or oriented toward conflict resolution had a negative impact on receiving messages. Having photos and being romantic and altruistic helped chances of success for women as well.

Also we observe that overall, the rate at which women receive messages is much more predictable than men judging by the R-squared for both models.

## *Homophily*

In order to answer RQ5 and RQ6 this paper uses a logistic regression to analyse whether homophily in sociodemographic or psychometric variables translates into higher chances of match communication. Homophily is operationalised as two users having the same value for any particular variable by creating a series of dummy variables. In the case of number of children and age however, it was more sensible to simply calculate the absolute value of the difference.[1] Whether two users communicate with each other was operationalised by creating two dependent variables: Communication and Initiation. Both are binary variables, which are set to one, whenever a suitable variable indicates that a user has replied to a message or initiated contact after a match by sending a message.

The logistic regression was run in a number of different specifications to hedge against omitted relevant variable bias and to test the robustness of the results. Variables were grouped by clusters into socioeconomic, personal, BAPIM, and Importance variables and regressions were run within clusters for all years, within clusters for each year separately, within each year with all clusters, and across all years with all clusters. The significance and sign of most coefficients varied severely, with the most rigorous specification rendering the majority insignificant. The only variables which seem to be fairly robustly significant and somewhat stable in their effects are *number of children*, the *desire for children* and a user's *smoking level*.

## Discussion and Conclusion

The development of evolutionary theories of human social behavior (Buss, 1989; Cunningham, 1986; Kenrick, Sadalla, Groth, & Trost, 1990; Symons, 1979; Thiessen & Gregg, 1980; Trivers, 1985) has afforded a strong theoretical framework for sex differences in mate selection criteria. The finding that women have consistently higher means across mate preferences in this work confirms findings of gender differences in mate preferences; namely that women are more selective and restrict their potential mating pool more than men do. (Hirsch et al. 2010) This finding has been found in literature about speed-dating as well (e.g., Fisman et al. 2006; Kurzban and Weeden 2005), and falls in line with theories in evolutionary biology about females being pickier about their potential mates. However, there are notable new findings in the work at hand that contradict previous investigations of mate preference in online daters.

For instance, Hirsch et al. claims that women have a stronger preference than men for income over physical attributes (Hirsch et. al, 2010). This work reveals that smoking level, ethnicity, and drinking level were the most important match criteria for both men and women overall, suggesting that ethnicity and lifestyle choices are important across both genders. In fact, income was the second least important criterion to women, religion being the least. Hirsch's claim is partially true, in that women on average do consider income in a potential match more important than men do, but the importance of this trait has decreased significantly over time. This change could theoretically be due to women's increased financial independence, though it would be difficult to attribute cause definitively.

The decline in importance of income, religion, and education for both men and women is a surprising trend that suggests perhaps people are becoming more tolerant and open to dating others outside of their own social strata. This tolerance has notably not translated over to age preferences, where patterns over time are less clear. Somewhat surprisingly, women are still more restrictive overall in their preference for age than men are. This may seem counterintuitive to those who might expect men to only seek mates within child-bearing age. As it turns out, women are pickier across the board, which may also have more to do with male over-representation on online dating sites and therefore increased female choice.

The finding that gender differences in response for the two lifestyle questions (smoking level and drinking level) were not significant from 2015 onward may reveal that social attitudes toward these activities is not gender-dependent. While the importance of drinking level for men rose from 2014-2017, both genders consistently regard drinking level "Somewhat Important," suggesting that social attitudes may have relaxed toward drinking level for both men and women. Meanwhile, preferences for smoking level became almost evenly split between those who consider smoking "Very Important" and "Not Important," suggesting that people in general fall into the two camps of smokers and non-smokers. Since smoking levels have decreased in the UK over time, this polarization of opinion may be due to changing demographics of the user base as well (UK Office for National Statistics, 2014).

---

[1] In another set of the analyses, instead of constructing dummy variables of the psychometric *BAPIM* scores and *Importance* variables, similarly the absolute value of the difference was taken. The latter reduces the risk of collinearity in the independent variable matrix and also allows for more flexibility in the model. Exact psychometric fits are likely to have a rare frequency and thus may impose too many restrictions on the model. Furthermore, treating a small difference in the psychometric scores as exactly the same as a large difference seems impractical and less realistic regarding the degree of compatibility between two users.

While women are more selective along virtually every mate preference criterion, this gender difference in selectivity crucially depends on group size. Previous literature has found that in smaller sessions (fewer than fifteen partners), selectivity is virtually identical for men and women, with subjects of each gender "saying yes" to about half of their partners. In larger group sizes, however, male selectivity is unchanged, while females become significantly more selective, choosing a little more than a third of their partners (Fisman et. Al, 2016). These results are quite distinct from the average difference in selectivity between men and women, suggesting rather more rapidly diminishing returns for increased dates for females when group size increases. Though Fisman's research focused on speed-dating, the parallel holds for the significantly increased group size of an online dating platform, where choice is virtually endless. The reasons women may be more selective than men and find less utility in increased choice could be manifold, from social stigma against women who go on many dates to differing motivations for why men and women use online dating in the first place.

The findings relating to the relationship between physical attractiveness and communication rate are notable due to their online context. Other studies attempting to measure the effects of physical attractiveness on popularity have encountered difficulties separating physical attractiveness from confounding characteristics, including social skills (Feingold, 1990; Goldman & Lewis, 1977). However, the online nature of this work is unique in that very little can be socially expressed from an online profile on eHarmony. After being matched through an algorithm, users are left to evaluate a profile based on little more than a picture.

The findings that physical attractiveness does not have a monotonous relationship with communication rate is somewhat surprising, but in line with previous research that produced similar findings (Fry, 2015). The slight but notable differences in the relationship for each gender have several implications. First, the higher rate of change for women scored between 2 and 6 in attractiveness suggests that women's communication rates are more dependent on their looks than for men. The finding that men value attractiveness more than women is consistent with previous research that found stronger correlations between opposite-sex romantic popularity for women than for men (Cohen, 1977). It is also in line with critical feminist theory as well as evolutionary and sociocultural theories of mate selection preferences that contend that men place greater value on physical attractiveness than do women. However, this is complicated as the rate of change is steeper and more negative for men ranked 6 and above. The same fear of rejection mentioned earlier may be stronger than for women initiating conversations with particularly attractive men.

The findings of growing asymmetry in communication initiation between men and women is rather counterintuitive. While early on, people might have hoped online dating would create a more equal playing field for women to initiate courtship, it has become clear that online dating has not only reflected but exacerbated male-dominated initiation. This is due largely to the lopsided activity levels for men and women on online dating sites, as women learn to expect male initiation and avoid initiation in keeping with learned norms. The introduction and mass popularity of mobile dating applications such as Tinder in 2014 could also explain the accelerated decline of female initiation over the following years, as online dating became more popular and the signaling and psychological costs for men sending messages declined.

As online dating becomes more popular and increasingly sophisticated, a new generation of dating apps are embedding costly signals into their platform design to solve this issue of lopsided communication. By instituting a mechanism whereby each agent has only a limited number of signals, they create opportunity costs associated with sending signals. For instance, Coffee Meets Bagel has a Woo button, where users pay (with the in-app "beans" currency) to send an extra signal to a specific someone. Users only get beans by performing tasks like inviting friends or purchasing them directly in the app store. In a similar vein, Tinder lets users send one Superlike per day. These signals work because they are costly to the sender by virtue of scarcity and the receiver knows this, so they pay attention to the signal in an otherwise noisy environment. The practical effects would be less messaging initiation from men but hopefully, more two-way communication for initiations that do take place.

Other factors that may influence the design of online dating platforms are our findings on individuals' 'dating capacity', or how frequently people engage with new prospective partners. Whilst user strategies might vary across more casual dating platforms, users on eharmony are particularly invested in finding a long term romantic partner, so we are confident that these findings are applicable to non-casual courtship behaviour in general. These findings and methods may nevertheless be integrated across dating platforms in order to enable effective platform-specific communication. Though research behind Dunbar's number acted as motivation for this line of inquiry, the picture is by no means complete. Specifically, further work is needed to assess the number of prospective partners that contact is simultaneously maintained with, rather than new people contacted, as well as cross-platform research to test results for serious communication patterns with different platform affordances, cultures, and sexualities. A related result for initiation rate (median ~1 per day) is provided in Fiore (2010), which is of comparable size to our own, though initiation is likely more sensitive to the platform population and design as compared to reply rate.

Our results across different dimensions of attractiveness by popularity and selectivity indicate individuals' awareness, if weak, of their own desirability. This, together with feedback from their level of success on the platform, informs user choices in initiating and replying. The correlations between attractiveness and selectivity for both the same and different modes of replying and initiating indicate that this awareness and strategy goes beyond directly addressing the individual games of active search for and deciding responses to potential partners.

From the models for the number of communications received, we learn that being younger and athletic and having more photos increases likelihood of receiving messages in online dating, as does being romantic and altruistic. These results add a more nuanced understanding to previous findings in RQ2 about the importance of physical attractiveness. It could be possible that being young and athletic is at least related to identifying as physically attractive, and that these traits increase likelihood of receiving messages. The negative relationship between communication rate and being older could suggest that age is not heterogeneous, but that users prefer younger potential partners. The negative relationship between communication rate and sending communication intiations also confirms that users who receive many communication intiations are less likely to send intiations themselves. It is unclear what signals users with higher levels of neuroticism and conflict resolution skills are sending in their profiles that decrease likelihood of receiving messages.

As for the differences between predicting success for men and women, the findings that drinking and being clever were positive predictors of success for men, but not for women were noteworthy. These findings suggest that physically reflected traits such as age and athleticism were most important factors for determining whether women would receive messages, in line with our earlier results in RQ2 about women being evaluated by their looks more than men.

The logistic regressions testing for the interplay of homophily and match communication uncover that a big difference in the number of children between two users seems to have strong negative effects on their chances of communicating or initiating contact. This seems intuitive, as the shared experience of having children or not having children has huge implications on the success of a potential relationship. Furthermore, the presence of children in a relationship introduces responsibility on a potential spouse that will greatly impact the seriousness of the relationship (s. Fiore et al. 2002).
Additionally, the desire for children is unsurprisingly decisive, with a matching desire leading to sometimes very strong positive effects. The choice of having or not having children is clearly a strong factor in a potential couple's life planning. Differing opinions on such life plans will clearly have a negative impact on a couple's shared experience.

However, the sometimes very strongly negative effect of a user's matching smoking level is rather interesting. Intuition seems to dictate that non-smokers will be very averse to the idea of entering a relationship with smokers, whereas smokers would either be indifferent or adverse to the idea of entering a relationship with a non-smoker. These results however indicate that matching smoking levels are detrimental to successful communication. One explanation might be that smokers do not want to enter a relationship with other smokers.

Smokers may be trying to quit smoking and being in close proximity to other smokers, or being emotionally involved with another smoker, strongly jeopardises this project[2].

The lack of robustly significant findings in the remainder of the variable set hints at low levels of homophily on online dating sites. Nor similarity in key socioeconomic, or psychometric variables seem to matter for the chances of successful match initiation or communication. Even when users seem to place nominal importance on variables such as income or ethnicity, they de facto appear to be insignificant.

This may be driven by a few factors: Firstly, men are much less selective in who they communicate with than women (s. above). Moreover, low activity of a large number of users, lower activity levels for women vis-à-vis men (Tong & Walter 2011), and the social and gender-normative conventions of men still typically being the first initiator for a conversation in dating settings (s. Rudder 2014, Finkel et al. 2012 Ansari and Klinenberg 2015, Hartford 2018) make it optimal for men to use the "shotgun method" of dating. That is, the optimal strategy is not to find a "good match" (i.e. high homophily levels), but to maximise the probability of a successful match by messaging a large number of people, irrespective of their potentially low fit. Secondly, it is very possible that people's reflections on what they find desirable in a potential match are overridden by more superficial concerns, such as a user's profile picture. Thirdly, it is possible that the relationship between homophily and successful communication is non-linear and thus is not captured by a logistic regression.

In conclusion, Our findings provide a quantitative overview of how heterosexual users seek mates, evaluate physical attractiveness, and communicate with one another in online dating systems. The results span investigations of various online dating phenomena first at the individual level, then between pairs of users, and finally between genders. In addition to its broader findings, this work sheds light on latent gender asymmetries in the user preferences and user behavior of online daters and how these differences have changed over time. The results from the aforementioned areas, and the regression of a number of them against the number of communications received on the platform, show that while there are many variations across mate preferences and communication patterns, there are few pointed variables that could actually act as potential predictors of sending and receiving messages overall.

Finally, one should note that this work was conducted within the context of data generated from one online dating site within the United Kingdom geography. While there was extra caution taken to ensure that findings were as representative as possible, it could always be valuable to confirm validity by testing novel data sets from other dating sites and from different geographies where social norms may differ. Future research could also strive to situate these findings within the offline courtship context. Building on this work, researchers could use a combination of qualitative and quantitative methods to better understand whether users approach mate preference, evaluation of physical attractiveness, and communication initiation differently in offline contexts where search costs or fear of rejection may be higher.

---

[2] It has to be acknowledged that in one regression the coefficient was positive, but in the majority of the regressions it was strongly negative. This result should be viewed carefully as some coefficients in the year-wise regression have suspiciously strong effects, pointing to a low robustness of the result.